\documentclass{article}[12pt]
\usepackage{hyperref}
\usepackage{amsmath,amssymb,latexsym,amsthm,mathtools}
\usepackage{url}
\usepackage{graphicx,color}
\usepackage{mathdots}
\usepackage{algorithm}
\usepackage{algorithmic}
\usepackage[margin=1.2in]{geometry}
\usepackage{caption}
\usepackage[utf8]{inputenc}
\usepackage{graphicx}
\usepackage{amsthm}
\usepackage{flushend} 
\usepackage{ulem}    
 \usepackage{float}
\usepackage{multirow}  
\usepackage{array}	   
\usepackage{booktabs} 
\usepackage{tabularx}  
\usepackage{longtable} 
\usepackage{tabu}      
\usepackage{threeparttable}

\DeclarePairedDelimiterX\braket[2]{\langle}{\rangle}{#1 \delimsize\vert #2}

\graphicspath{{figures/}}

\title{Modeling Population Human Mobility with Dynamic Mode Decomposition}
\author{Liantao Li and Yang Yang}
\date{}
\begin{document}
\maketitle

\abstract{
Human mobility research concerns spatiotemporal individual and population movement. Accurate modeling and prediction of human mobility can provide opportunities to monitor, manage and optimize human movement for improved social-economic benefit. In this paper, we adopt the dynamic mode decomposition algorithm to model population human mobility using visitor flow data between different states in the United States from 2019 to 2021~\cite{kang2020multiscale}. We train multiple DMD models with different low rank structures, and evaluate their modeling accuracy and predictability on novel testing data.
}

\section{Introduction}

Human mobility research studies individual and collective human movement in space and time.
Examples of human mobility include livelihood mobility, urban migrations, and traffic flows.
Understanding human mobility is essential to multiple social-economic aspects such as mitigating urban traffic congestion and improving public health~\cite{barbosa2018human}. For instance, human mobility plays a significant role in understanding the dispersal of infectious diseases such as COVID-19. Comprehension of human mobility could help predict the variation of diseases to enable optimal decision making and public health interventions at early stages~\cite{zheng2021exploring}.


\medskip
\noindent An important aspect of human mobility study is modeling and prediction of human mobility patterns. Accurate modeling and prediction can provide opportunities to monitor, manage and optimize human movement. For example, effective modeling of urban traffic flow could suggest strategies to mitigate traffic jam, promoting road safety and reducing environmental impacts. Human mobility at different scales exhibit different natures. Mobility on the individual scale typically depends on a large number of uncertainty factors such as individual mood or emotions; it thus often shows a stochastic nature. On the other hand, human mobility on the scale of groups demonstrates a deterministic nature, and the nature can be understood as an average/expectation of a large number of individual mobility.

\medskip
\noindent Based on the scale of mobility, current models can be classified into two categories: individual mobility model and general population model. 
Individual models aims at reproducing individual mobility patterns. Typical examples include Brownian motion model, Levy flight model, continuous time random walk model, and preferential return model. In contrast, general population models seeks to describe the aggregate mobility of many individuals. 
Representatives include the gravity model proposed by Zipf, which was inspired by Newton's Law of Gravitation~\cite{zipf1946p}; intervening opportunities models introduced by Stouffer~\cite{stouffer1940intervening}, and the radiation model proposed by Simini et al.~\cite{simini2012universal}, which is parameter-free and can be used without mobility measurement. We refer readers to the reference~\cite{barbosa2018human} for a comprehensive review of different models as well as their weakness and strength.

\medskip
\noindent As the average of numerous individual behaviors, mobility of populations usually features significantly higher stability and predictability. We therefore focus on modeling population mobility in this paper.
All the aforementioned population mobility models are based on the intrinsic physics, thus depending on accurate prior knowledge to achieve accurate modeling.
On the other hand, the development of information and communication technologies
has made large-scale mobility-related datasets more accessible than ever before.
With the advance of machine learning methodology, powerful data-driven machinery becomes available to extract statistically meaningful patterns from these large datasets. The goal of this paper is to evaluate a data-driven, equation-free modeling approach known as the dynamic mode decomposition.

\medskip
\noindent The dynamic mode decomposition(DMD) is a data-driven method that can provide effective modeling of dynamical systems~\cite{kutz2016dynamic}. The method was initially proposed by Schmid and Sesterhenn~\cite{schmidandsesterhenn} and Schmid~\cite{schmid2010dynamic} 
in fluid dynamics, yet later found wide applications in a variety of areas including video processing~\cite{grosek2014dynamic}, epidemiology~\cite{proctor2015discovering}, neuro-science~\cite{brunton2016extracting}, and financial trading~\cite{mann2016dynamic}.
The idea of DMD is to learn a temporal low rank structure of the dynamic systems. It is therefore also a dimension reduction technique, which makes it ideal for processing large-scale data.
In this paper, we will apply DMD to model the population human mobility between different states in the US from 2019 to 2021 using the publicly available dataset~\cite{kang2020multiscale}. We will build multiple DMD models with different low rank structures, and evaluate their accuracy in predicting future human mobility patterns.

\medskip
\noindent The paper is organized as follows. We describe the structure of the human mobility data set in use in Section 2. Section 3 presents the idea of DMD in detail and formulates the precise algorithm. Section 4 includes multiple numerical experiments to validate the efficacy of DMD in modeling population human mobility with different training and testing datasets.

\section{Data Structure}


The dataset we will utilize to model human mobility is the multis-cale dynamic human mobility flow dataset collected in~\cite{kang2020multiscale} and made publicly available on Github~\cite{github}.
It consists of daily and weekly human mobility flow data collected across the United States from 2019 to 2021.
The dataset is collected with the help of the mobile phone application ``SafeGraph'' and pre-processed in several steps. First, selected mobile phone users from an origin to various destinations are tracked to generate visit information. Next, users in selected regions are grouped to form multiple clusters, and daily/weekly visit information between two clusters are calculated by summing up the number of individual users.
Next, data are aggregated into three spatial scales: census tract, county and state by grouping all records based on the origin-destination(O-D) pairs. Finally, the dynamic population flow from the geographic origin ``$o$'' to the geographic destination ``$d$'' is inferred from the visitor flow using the following relation:
\begin{equation} \label{eq:popflow}
pop\_flows(o,d)= visitor\_flows(o,d)\times \frac{pop(o)}{num\_devices(o)}
	\end{equation}
where $pop\_flows(o,d)$ is the estimated dynamic population flows from the origin $o$ to the destination $d$, $visitor\_flows(o,d) $ is the computed mobile phone-based visitor flow from $o$ to $d$, $pop(o)$ is the population
at the origin $o$ extracted from the American Community Survey~\cite{webster2006us}, and $ num\_devices(o)$ is the number of unique mobile devices residing in $o$. Details description of these steps are described in~\cite{kang2020multiscale}.


\medskip
\noindent In this paper, we will make use of weekly visit flow data between different states. The data includes 52 distinct places as origins and destinations: 50 states, Washington D.C., and Puerto Rico.
The weekly visitor flow data is recorded in a data matrix~\cite{github} with nine columns of information consisting of unique identifiers for origins and destinations, geographic locations of origins and destinations, data range of the records, estimated visitor flow, and inferred population flow based on the relation~\eqref{eq:popflow}.
The information that is directly used in this paper is identifiers for origins and destinations, data range, and estimated visitor flow. We attempt to utilize the information to model and predict collective human mobility patterns.

\section{DMD Algorithm}
The dynamic mode decomposition(DMD) is an equation-free and data-driven method in modeling dynamic systems using high dimensional data~\cite{kutz2016dynamic,Tu2014OnDM,bistrian2015improved,brunton2016extracting,mohan2018data}. It is designed to learn coherent patterns of dynamic systems.
Specifically, for a dynamical system
\begin{equation} \label{eq:ds}       
     \frac{dx}{dt} = f(x,t;\mu)        
	\end{equation}
where $x(t)\in \mathbb{R}^n$ is the state of the system at time $t$, $\mu$ represents parameters of the system, and $f(x,t;\mu)$ represents the dynamics, its corresponding discrete-time form can be obtained by evolving~\eqref{eq:ds} by a fixed time step $\Delta t$. Denote $x_k := x(k\Delta t)$ where $x$ is the solution of~\eqref{eq:ds}, then the Fundamental Theorem of Calculus implies
\begin{equation}       
   x_{k+1} = x_k+ \int_{k\Delta t}^{(k+1)\Delta t}f(x,\tau;\mu)d\tau
\end{equation}
It is natural to introduce the following discrete-time flow map $F$:
\begin{equation}       
    x_{k+1} := F(x_k) := x_k+ \int_{k\Delta t}^{(k+1)\Delta t}f(x,\tau;\mu)d\tau.
\end{equation}

\noindent Given snapshots of the solution $x_1, x_2, \dots, x_{m}$, one can use these snapshots to  construct two $n\times (m-1)$ data matrices:
\begin{equation}       
		X=\left(                
		\begin{array}{cccc}   
			| & | &  & |\\  
			x_1 & x_2 & ... &x_{m-1}\\  
			| & | &  &|\\
		\end{array}
		\right), \quad\quad\quad 
		X'=\left(                
		\begin{array}{cccc}   
			| & | &  & |\\  
			x_2 & x_3  & ... &x_m \\  
			| & | &  &|\\
		\end{array}
		\right).
\end{equation}
The idea of DMD is to compute the leading eigendecomposition of the best-fit linear operator $A$ relating the data $X^{'} \approx AX$:
\begin{equation}       
	A=X'X^{\dagger}            
\end{equation}
where $X^{\dagger} $ is the pseudoinverse of $X$. The eigenvectors of $A$ are said to be the DMD modes. More specifically, the DMD algorithm~\cite{kutz2016dynamic} is summarized in Algorithm~\ref{alg:DMD}:

\begin{algorithm}
\caption{DMD Algorithm~\cite{kutz2016dynamic}}\label{alg:DMD}
   1. Arrange the data $\{x_1,...,x_{m}\}$ into matrices 
     \begin{equation}       
		X = \left(                
		\begin{array}{cccc}   
			| & | &  & |\\  
			x_1 & x_2 & ... &x_{m-1}\\  
			| & | &  &|\\
		\end{array}
		\right)    ,          
		X' =\left(                
		\begin{array}{cccc}   
			| & | &  & |\\  
			x_2 & x_3  & ... &x_m \\  
			| & | &  &|\\
		\end{array}
		\right) 
	\end{equation}
where  $X$ and $X'$ are defined in equation (5) and equation (6) respectively.\\
\\
	 2. Compute the (reduced) SVD of $X$, writing
	 \begin{equation}       
	 X = U\Sigma V^*
	\end{equation}
	\indent where $U$ is an $n\times r$ matrix, $\Sigma$ is an $r\times r$ diagonal matrix, $V$ is an $m\times r$ matrix, and $r$ is the rank of $X$ (known as the \textit{target rank}.).\\
	\\
	 3. Define the matrix
	 \begin{equation}       
	 \widetilde{A} \vcentcolon= U^*X'V\Sigma^{-1}
	\end{equation}
	\\
	4. Compute the eigendecomposition of $ \widetilde{A}$, writing
	 \begin{equation}       
	 \widetilde{A}W = W\Lambda
	\end{equation}
	\indent where columns of $W$ are eigenvectors and $\lambda$ is a diagonal matrix containing 
	\indent the corresponding eigenvalues $\lambda_k$.\\
	\\
	 5. Reconstruct the eigendecomposition of $A$ from $W$ and $\Lambda$. The eigenvalues 
	\indent of A are given by $\Lambda$ and the eigenvectors of $A$ (DMD modes) are given by 
	\indent columns of $\Phi$:
	\begin{equation}       
	 \Phi = X'V\Sigma^{-1}W
	\end{equation}
	\\
	 6. Finally, the approximate solution at all future
times is given by
\begin{equation} \label{eq:xt} 
x(t)\approx \sum_{k=1}^{r} \phi_k \exp(\omega_k(t)) b_k =\Phi  \exp(\Omega t) \textbf{b},
\end{equation}
\indent where $\omega_k = ln(\lambda_k)/\Delta t $ , $b_k$ is the initial amplitude of each mode, $\Phi$ is the 
\indent matrix whose columns are the DMD eigenvectors $\phi_k$, and $\Omega = diag(\omega)$ is a 
\indent diagonal matrix whose entries are the
eigenvalues $\omega_k$. The eigenvectors $\Phi$ 
\indent are the same size as the state $x$, and $\textbf{b}$ is a vector of the coefficients $b_k$.
\end{algorithm}

\section{Numerical Experiments}

We will apply the DMD algorithm to the weekly visit flow data~\cite{github} to model human mobility patterns. The 52 distinct places (50 states plus Washington D.C. and Puerto Rico) are assigned unique identifiers from 1 to 52.
The weeks are arranged in the chronological order, starting from the first week of 2019.
For the $t$-th week, we form a $52\times 52$ matrix $A^t$ whose $(i,j)$-entry is the visitor flow data from the origin $i$ to the destination $j$. 
Next, we symmetrize each matrix $A^t$ to obtain a matrix $S^t$ whose $(i,j)$-entry is the total visitor flow from $i$ to $j$ or vice versa. If we view the matrix $A^t$ as the adjacent matrix of a directed graph defined by the visitor flow, then $S^t$ is the adjacent matrix of the underlying undirected graph. We then re-shape each matrix $S^t$ to a column vector and use it as a temporal snapshot of the underlying dynamic system that dictates human mobility to apply the DMD algorithm. 
By the end of these procedures, we obtain a data matrix 
 \begin{equation}  \label{eq:Mmatrix}
		M = \left(                
		\begin{array}{cccc}   
			| & | &  & |\\  
			S^1 & S^2 & ... &S^{m}\\  
			| & | &  &|\\
		\end{array}
		\right)   
	\end{equation}
where the columns are reshaped symmetric matrices $S^t$, $t=1,2,\dots,m$.
The procedures are summarized in Algorithm~\ref{alg:Data Matrix M}.

\begin{algorithm}
\caption{Generate Data Matrix M}\label{alg:Data Matrix M}
   \begin{algorithmic}
	    \FOR{$t=1, \dots, m$}
	         \STATE
                \begin{enumerate}
    		    \item\textbf{Create Origin-Destination Matrix $A_{k\times k}^t$ for the t-th week} \\
  \begin{equation}       
 	A^t_{ij} = visitor\_flows_t(i,j)  
 \end{equation}
where $visitor\_flows_t(i,j)$ is the t-th week visitor flow
 from state i to state j and there are k states in total.
\item \textbf{Get Symmetric Matrix $S_{k\times k}^t$ }\\
   \begin{equation}       
 	S^t = (A^t+{A^t}^T)/2
 \end{equation}
		    \item \textbf{Reshape Matrix $S_{k\times k}^t$ and Fill Data Matrix $M_{k^2\times m}$ }\\
Finally, we convert square matrix $S_{k\times k}^t$ to a  $k^2$-dimentional column vector $M_t$ by column  and 
assign the vector $M_t$  to t-th column of matrix M.
                \end{enumerate} 
	    \ENDFOR
	\end{algorithmic}
\end{algorithm}
\subsection{Experiment 1.}
In the next few subsections, we conduct numerical experiments by applying Algorithm~\ref{alg:Data Matrix M} to model the human mobility patterns.

\begin{figure}[H]
    \centering 
    \includegraphics[width=0.6\textwidth]{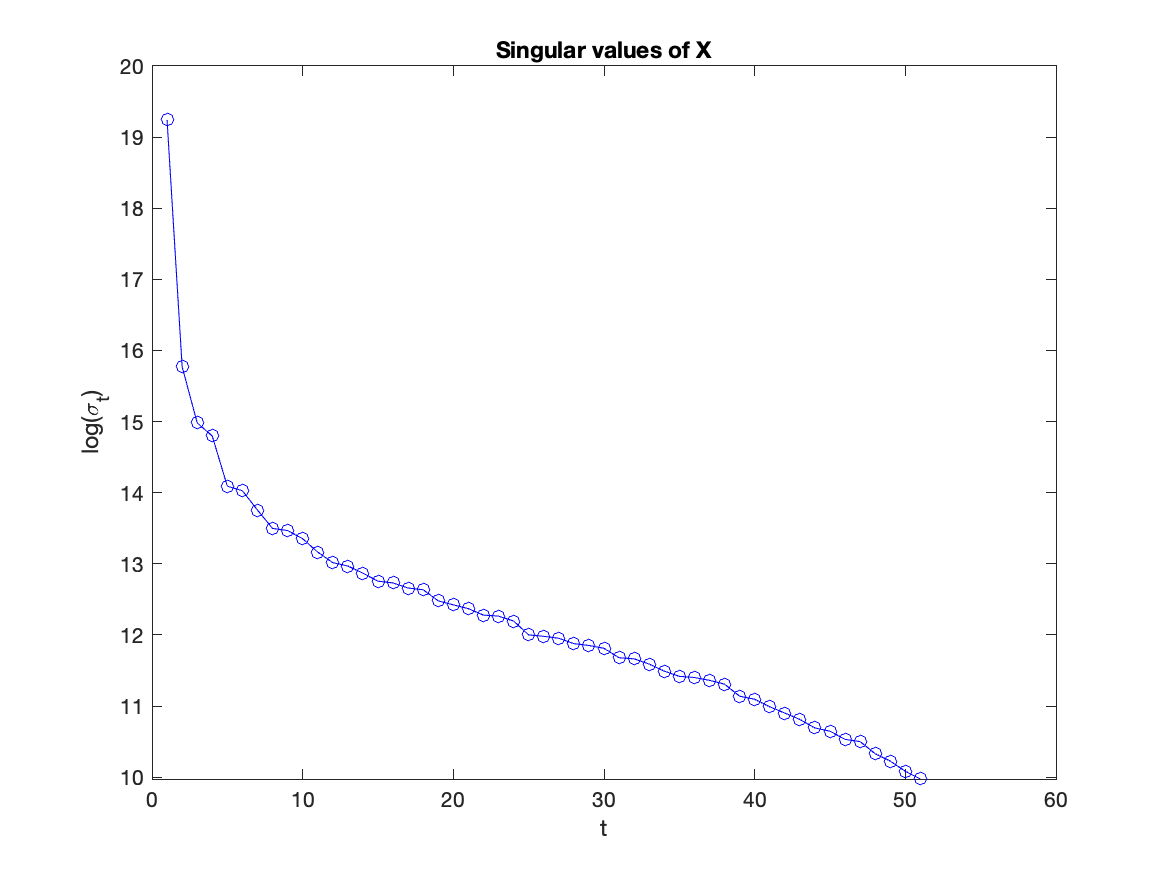} 
    \caption{Singular Value of 2019 Data Matrix X}
    \label{Fig.Singular Value of 2019}
\end{figure}

\medskip
\noindent Firstly, we use the visitor flow data of the year 2019 (52 weeks in total). We implement Algorithm~\ref{alg:DMD} to obtain the DMD model $x(t)$ as defined in~\eqref{eq:xt}.
The singular values of the matrix $M$ are plot in Fig.~\ref{Fig.Singular Value of 2019} in log scale. The distribution decreases slower after the 10th singular value. 
The DMD models~\eqref{eq:xt} with target rank $r=5,10,15$ are calculated following Algorithm~\ref{alg:DMD}.

\begin{figure}[H]
    \centering
    \includegraphics[width=0.6\textwidth]{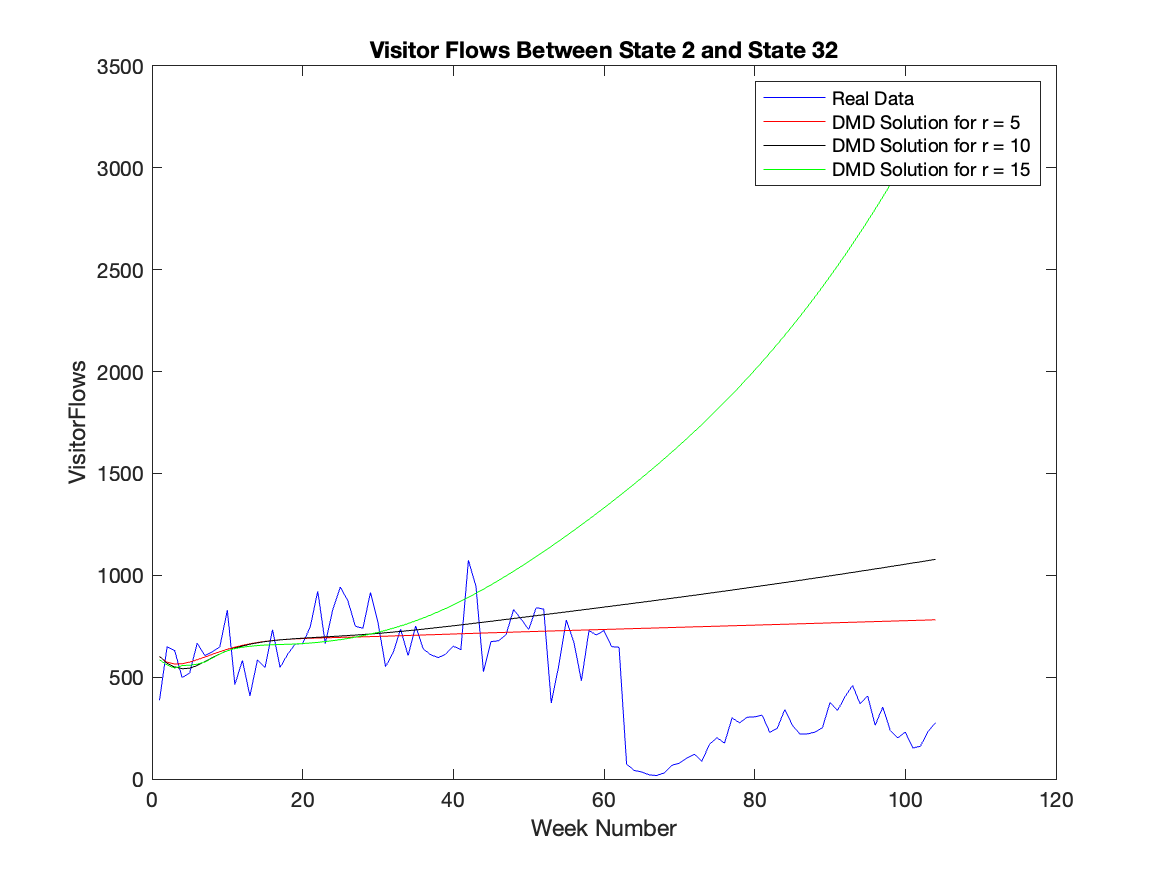}
    \caption{Real Visitor Flow and DMD Predictions in 2019 and 2020}
    \label{Fig.Visitor Flow Comparison From 2019 to 2020}
\end{figure}

\medskip
\noindent To evaluate the model accuracy, we randomly picked the origin with ID=2 and the destination with ID=32. The visitor flow data as well as DMD models between these two places from Week 1 to Week 105 (that is, from 2019 to 2020) are illustrated in Fig.~\ref{Fig.Visitor Flow Comparison From 2019 to 2020}. We observe that the DMD models with $r=5,10$ remains to fit well on the testing data from Week 57 to Week 62 (that is, the first 5 to 10 weeks in 2020), yet the DMD model with $r=15$ flows up more rapidly than others. On the other hand, all the models seem to deviate from the real data after Week 62 (that is the 10th week in 2020).

\medskip
\noindent Various errors between the real data and DMD model predictions with target rank $r=10$ are shown in Table~\ref{Table.Error results in 2020}. Here $M$ refers to the data matrix~\eqref{eq:Mmatrix}, $M_{dmd}$ consists of predictions computed by~\eqref{eq:xt}, $||M_i||_2$ is the $L^2$ norm of the visitor flow of Week $i$, $||M_{dmdi}||_2$ is the the $L^2$ norm of the DMD prediction of Week $i$.
$\frac{||M_i-M_{dmdi}||_2}{||M_i||_2}$ and $\frac{||M_i-M_{dmdi}||_{\infty}}{||M_i||_\infty}$ are the relative $L^2$ and $L^{\infty}$ prediction error, respectively. It is clear that, DMD predictions can be far away from real data in the long run, as the model~\eqref{eq:xt} grows exponentially in time $t$.

\subsection{Experiment 2.}
\begin{figure}[H]
    \centering
    \includegraphics[width=0.6\textwidth]{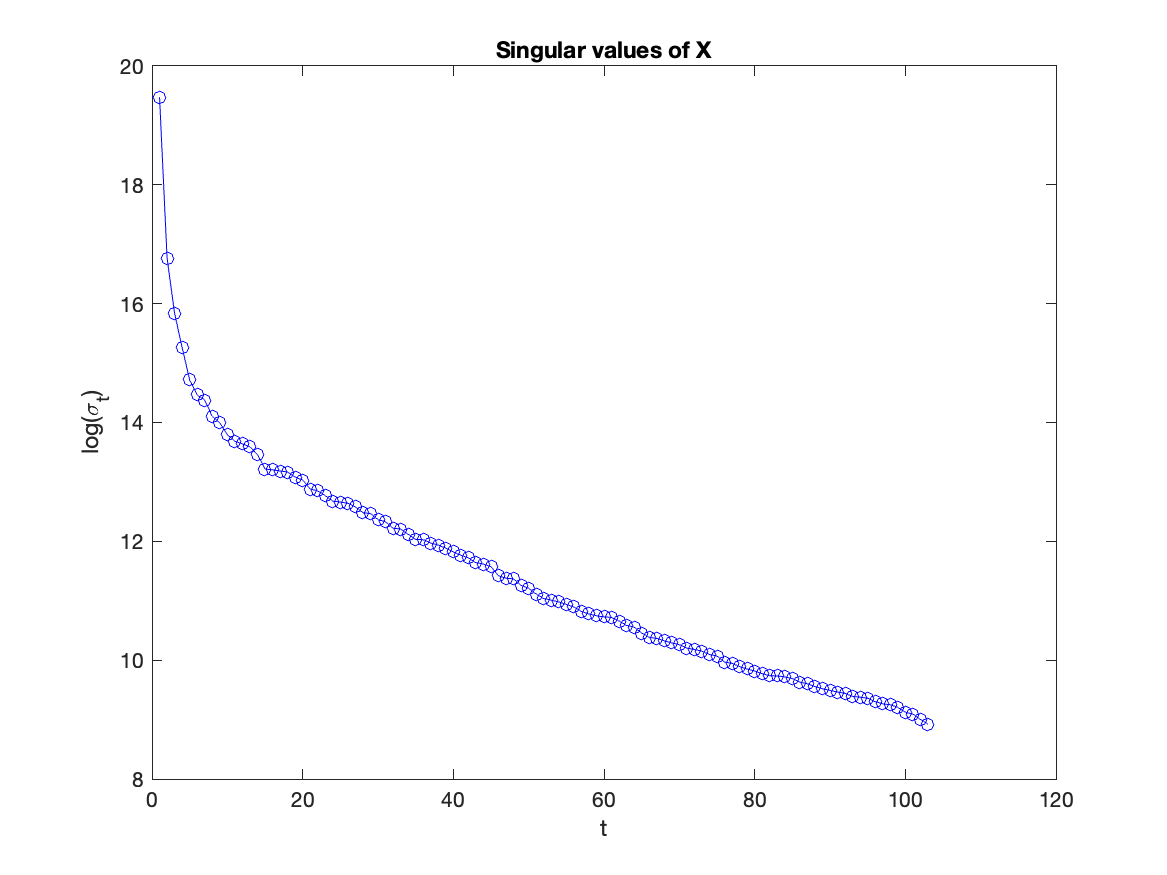}
    \caption{Singular Value of 2019 and 2020 Data Matrix X}
    \label{Fig.Singular Value of 2019 and 2020}
\end{figure}

\begin{figure}[H]
    \centering
    \includegraphics[width=0.6\textwidth]{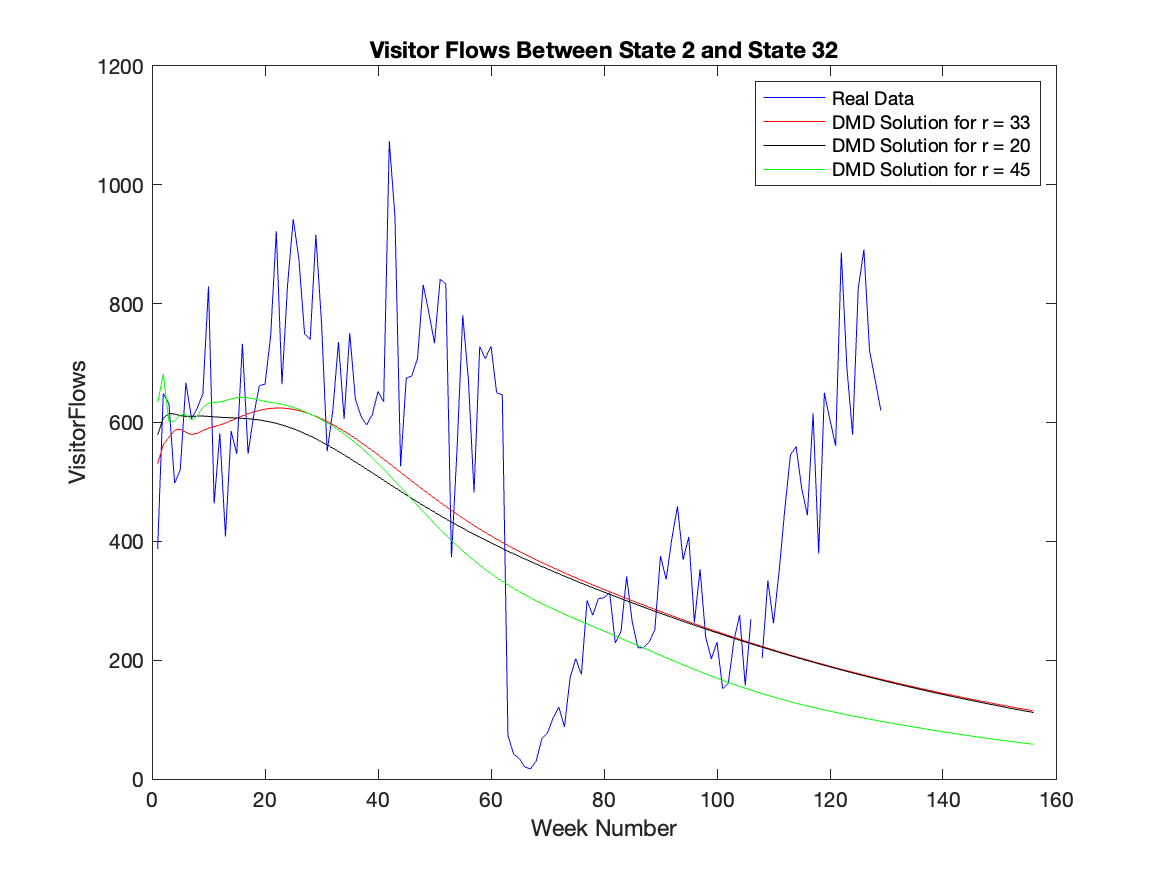}
    \caption{Real Visitor Flow and DMD Computation Visitor Flow Data from 2019 to 2021}
    \label{Fig.Real and Numerical Data Comparison from 2019 to 2021}
\end{figure}

\noindent In this experiment, we use 2019 and 2020 data (104 weeks in total) to compute the DMD mode to predict 2021 data.
The singular values of the data matrix $M$ is shown in Fig.~\ref{Fig.Singular Value of 2019 and 2020}. We will choose target rank $r=20,33,45$ to compute different DMD models. We choose the places with ID=2 and ID=32 again. The real visitor flow data and DMD predictions from the first week of 2019 to the last week of 2021 (156 weeks in total) are shown in Fig.~\ref{Fig.Real and Numerical Data Comparison from 2019 to 2021}. Various errors with rank $r=33$ are summarized in Table~\ref{Table.The Error Results in 2021}. This time, none of the DMD models seems to give reasonable predictions on testing data.



\subsection{Experiment 3.}
\begin{figure}[H]
    \centering
    \includegraphics[width=0.6\textwidth]{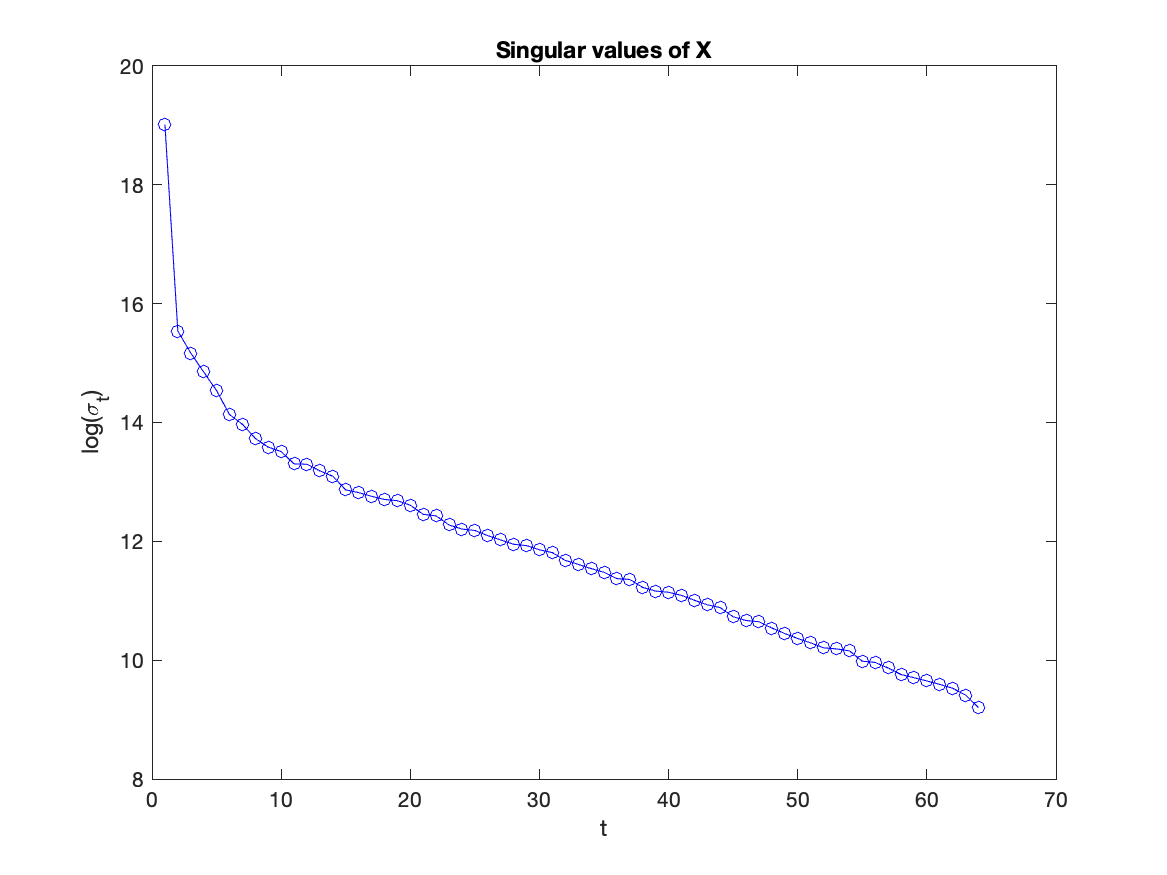}
    \caption{Singular Value of Partly 2019 and 2020 Data Matrix X}
    \label{Fig.Singular Value of Part 2019 and 2020}
\end{figure}

\begin{figure}[H]
    \centering
    \includegraphics[width=0.6\textwidth]{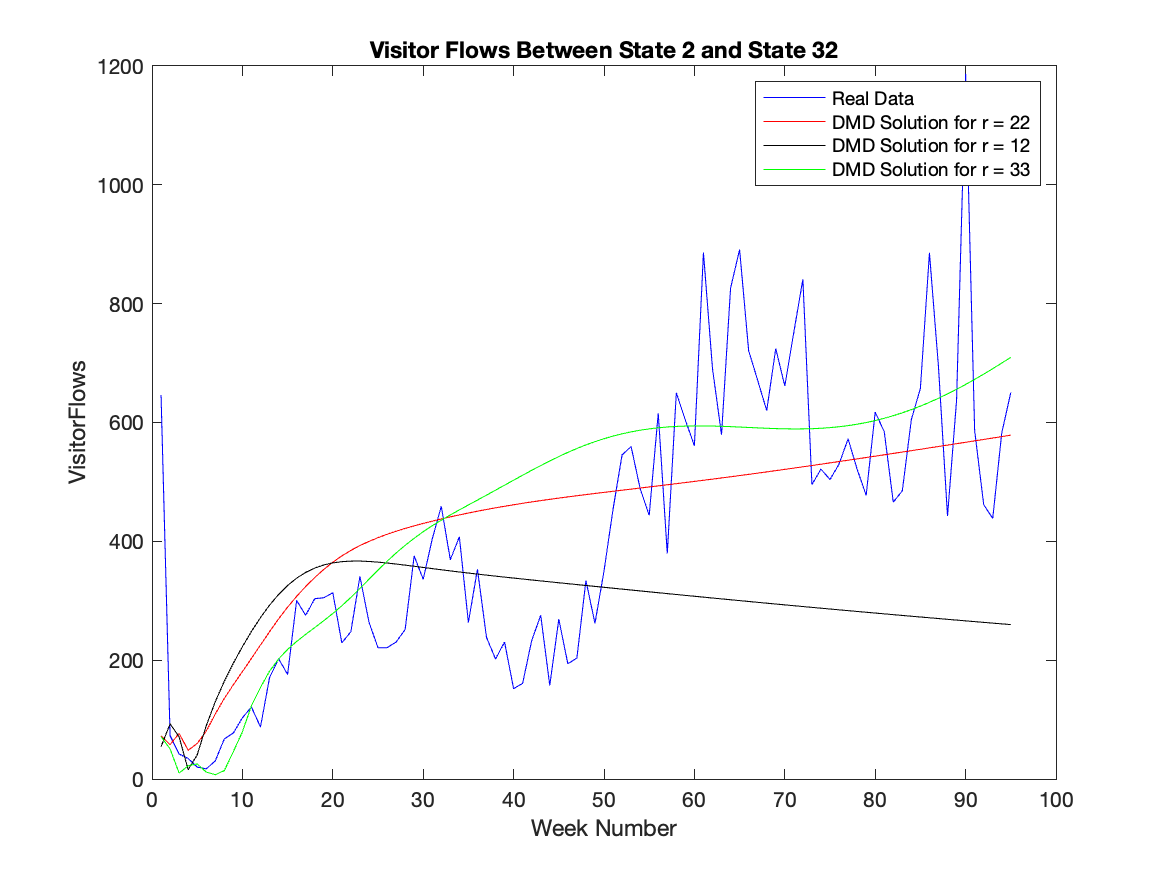}
    \caption{Real Visitor Flow and DMD Computation Visitor Flow Data From March 9, 2020 to December 31, 2021}
    \label{Fig.VisitorFlowComparison}
\end{figure}

We repeat the experiment using visitor flow data from March 9,2020 to May 31,2021 (65 weeks in total) as the training data to compute DMD models. The singular values of the data matrix are shown in Fig.~\ref{Fig.Singular Value of Part 2019 and 2020}. The real data and DMD predictions with target rank $r=12,22,33$ are displayed in Fig.~\ref{Fig.VisitorFlowComparison}, and the errors are summarized in Table~\ref{table:errorin2021}.

\medskip
\noindent We observe that the predictions are better than those in Experiment 2. The difference between the training data for Experiment 1 \& 3 and that for Experiment 2 is that the latter contains a sharp change in March 2020. This was because of the presidential declaration of the state of national emergency on March 13, 2020, which leads to a drastic decay of the number of domestic visitors. The DMD models, as is shown in~\eqref{eq:xt}, consist of mostly low-frequency content, thus finds it difficult to learn this high-frequency content.


\newpage
\graphicspath{{Figure/}}

\graphicspath{{Figure/}}

\begin{table}[H]
	\begin{center}
		\caption{The Weekly Visitor Flows Data and Errors in 2020 }
			\label{Table.Error results in 2020}
		\begin{tabular}{cccccc}
			\toprule
			Week of 2020 & $||M_i||_2$  & $||M_{dmdi}||_2$ & $ \frac{||M_i-M_{dmdi}||_2}{||M_i||_2}$ & $ \frac{||M_i-M_{dmdi}||_{\infty}}{||M_i||_\infty}$ \\
			\midrule
			1st	&2.2066e+07 &2.0900e+07  &0.0776 &0.0508	\\
			2nd &  2.1741e+07&  2.2276e+07	&0.0679&0.0694	\\
			3rd	&  2.1410e+07 & 2.2861e+07  &0.0669&0.0732	\\
			4th	& 2.1071e+07 &  2.2923e+07  &0.0713&0.0742	\\
			5th	&   2.2923e+07&  2.2719e+07&0.0776&0.0558	\\
			\bottomrule
		\end{tabular}
	\end{center}
\end{table}

\newpage

\graphicspath{{Figure/}}

\graphicspath{{Figure/}}

\begin{table}[H]
	\begin{center}
		\caption{The Weekly Visitor Flows Data and Errors in 2021 }
			\label{Table.The Error Results in 2021}
		\begin{tabular}{cccccc}
			\toprule
			Week of 2021 & $||X_i||_2$  & $||M_{dmdi}||_2$ & $ \frac{||M_i-M_{dmdi}||_2}{||M_i||_2}$ & $ \frac{||M_i-M_{dmdi}||_{\infty}}{||M_i||_\infty}$ \\
			\midrule
			1st	&2.1651e+07 &1.9726e+07  &0.1040 &0.1088	\\
			2nd &  2.1370e+07& 1.9628e+07	&0.0958&0.1073	\\
			3rd	&  2.0825e+07 & 1.9530e+07  &0.0788&0.0782	\\
			4th	& 2.0556e+07&  1.9432e+07  &0.0779&0.0901	\\
			5th	&   2.1252e+07&  1.9335e+07&0.1074&0.1192	\\
			\bottomrule
		\end{tabular}
	\end{center}
\end{table}

\newpage 
\graphicspath{{Figure/}}
\graphicspath{{Figure/}}

\begin{table}[H] 	
	\begin{center}
		\caption{The Weekly Visitor Flows Data and Errors in 2021 }
			\label{table:errorin2021}
		\begin{tabular}{cccccc}
			\toprule
			Week of 2021 & $||X_i||_2$  & $||M_{dmdi}||_2$ & $ \frac{||M_i-M_{dmdi}||_2}{||M_i||_2}$ & $ \frac{||M_i-M_{dmdi}||_{\infty}}{||M_i||_\infty}$ \\
			\midrule
			23rd	&2.6570e+07 &2.8383e+07  &0.0733 &0.0576	\\
			24th &  2.6968e+07& 2.8500e+07	&0.0640&0.0633	\\
			25th	& 2.6561e+07 & 2.8617e+07  &0.0841&0.0754	\\
			26th	& 2.6720e+07&   2.8735e+07  &0.0824&0.0703	\\
			\bottomrule
		\end{tabular}
	\end{center}
\end{table}

\section*{Acknowledgement}
The research of L. Li and Y. Yang is partially supported by the NSF grant DMS-1715178, DMS-2006881, and the start-up fund from Michigan State University.

\normalem
\bibliography{refs}
\end{document}